\documentclass[modern,twocolumn]{aastex62}
\usepackage{float}

\def\M{M$_{\odot}$}
\def\Z{Z$_{\odot}$}
\def\ni{$^{56}$Ni}
\def\fe{\ion{Fe}{2}\,$\lambda$5018}

\shorttitle{Bright Type IIP SNe in Low-Metallicity Galaxies}
\shortauthors{Scott et al.}

\begin{document}

\title{Bright Type IIP Supernovae in Low-Metallicity Galaxies}

\correspondingauthor{Spencer Scott}
\email{spencerscott@college.harvard.edu}

\author[0000-0001-7705-7388]{Spencer Scott}
\affil{Center for Astrophysics \textbar Harvard \& Smithsonian, 60 Garden Street, Cambridge, MA, 02140, USA}

\author[0000-0002-2555-3192]{Matt Nicholl}
\affiliation{Institute for Astronomy, University of Edinburgh, Royal Observatory, Blackford Hill, Edinburgh EH9 3HJ, UK}
\affiliation{Birmingham Institute for Gravitational Wave Astronomy and School of Physics and Astronomy, University of Birmingham, Birmingham B15 2TT, UK}

\author{Peter Blanchard}
\affiliation{Center for Astrophysics \textbar Harvard \& Smithsonian, 60 Garden Street, Cambridge, MA, 02140, USA}

\author[0000-0001-6395-6702]{Sebastian Gomez}
\affiliation{Center for Astrophysics \textbar Harvard \& Smithsonian, 60 Garden Street, Cambridge, MA, 02140, USA}

\author{Edo Berger}
\affiliation{Center for Astrophysics \textbar Harvard \& Smithsonian, 60 Garden Street, Cambridge, MA, 02140, USA}

\begin{abstract}
	We present measurements of the pseudo-equivalent width of the \fe\ absorption feature in the spectra of 12 Type IIP supernovae (SNe II) in low luminosity ($M>-17$) dwarf host galaxies. 
    The \fe\ line has been suggested to be a useful diagnostic of the metallicity of the SN progenitor stars.
	The events in our sample were discovered photometrically by the Pan-STARRS Survey for Transients (PSST), and classified spectroscopically by us.
	Comparing our sample to 24 literature SNe II, we find that those in low luminosity hosts have significantly weaker \ion{Fe}{2} features, with a probability of $10^{-4}$ that the two samples are drawn from the same distribution.
    Since low-mass galaxies are expected to contain a lower fraction of metals, our findings are consistent with a metallicity dependence for \fe, and therefore support the use of this line as a metallicity probe, in agreement with a number of recent works.
	In addition, we find that the SNe in faint (low-metallicity) hosts may be more luminous on average than those in the literature sample, suggesting possible physical differences between Type IIP SNe at low and high metallicity. However, accurate determinations of host galaxy extinction will be needed to quantify such an effect.
\end{abstract}

\keywords{supernovae:general}

\section{Introduction} \label{sec:intro}

Most stars with initial masses $\gtrsim 8$\,\M\ end their lives in supernova (SN) explosions when their iron cores collapse. The spectroscopic properties of SNe depend on the distribution of elements in their progenitor stars, with the most common type, Type II SNe (or SNe II), being defined by strong and broad Balmer lines -- thus these stars have retained their hydrogen envelopes until the moment of explosion. Serendipitous detections of the progenitors in pre-explosion images have confirmed that SNe II result from red supergiants \citep{sma2009}.

Historically, SNe II have been divided into sub-classes based on the morphology of their light curves: either a $\sim 100$ day `plateau' in luminosity (SN IIP), or a `linear' decline (SN IIL). More detailed studies from a range of authors have revealed a richer picture, with a range of plateau durations, decline rates, and luminosities, and evidence for a more continuous distribution of properties \citep[e.g.][]{arcavi2012,anderson2014,faran2014,sanders2015,gall2015,gonzalez2015,pejcha2015,valenti2016,rubin2016,gutierrez2017}. Along with progenitor radii and masses, and synthesised \ni\ mass, the presence of circumstellar material lost prior to explosion is likely responsible for some of this diversity. All of these factors can be influenced by the metallicity of the progenitor star. Therefore constraining the metallicity of SNe II is key to fully understanding their properties.

\citet{dessart2014} showed using spectral models that the equivalent width of the \fe\ absorption line can be a useful proxy for metallicity in SNe II, with higher metallicity resulting in an equivalent width of larger absolute value at a given time from explosion. This was confirmed observationally by \citet{anderson2016}, who compared results from a large sample of SN II spectra to the metallicities indicated by their host galaxy spectra. As this sample was compiled from the literature, it consisted primarily of SNe II in massive host galaxies, which were all at metallicities greater than that found in the Large Magellanic Cloud. 

\citet{taddia2016a} presented an untargeted sample of SNe II from the Palomar Transient Factory, spanning a wider range of host environments. They found several events in faint galaxies, with lower \fe\ equivalent widths than those in the sample of \citet{anderson2016}. Fainter galaxies typically have lower metal fractions the 'mass-metallicity relation'; \citep[e.g.][]{tremonti2004, kewley2008}, and indeed some events from \citet{taddia2016a} were consistent with $Z \sim 0.1$\,\Z\ \citep{dessart2014}.

The photometric properties of SNe II also seem to be different at low-metallicity. \citet{taddia2016a} noted a possible correlation suggesting lower-metallicity SNe II were more luminous. More recently, \citet{gutierrez2018} conducted a detailed statistical study analysing a sample of SNe II selected in faint galaxies, comparing to the higher-metallicity samples of \citet{anderson2016} and \citet{gutierrez2017}. As well as confirming that the SNe II in low-luminosity (metal-poor) galaxies have weaker \fe\ absorption features, they also found that these events tended to have light curves with slower decline rates on the plateau.

In this letter, we present and analyse a new sample of SNe II in faint galaxies. We measure the strength of their \fe\ absorption lines and estimate their plateau luminosities, and compare these properties as well as their host galaxy magnitudes to a literature sample of SNe II, primarily in massive galaxies. We show that our SNe have weaker \fe\ absorption, confirming results from \citet{dessart2014,anderson2016,taddia2016a,gutierrez2018}. We find that our sample of low-metallicity SNe II are significantly brighter, by up to $\sim 1$\,mag, than the literature sample, suggesting metallicity is an important factor in SN II evolution.

\section{Methods} \label{sec:methods}

\subsection{Sample of SNe II in faint galaxies}
\label{sec:select}

Our SNe were selected from the Pan-STARRS Survey for Transients (PSST; \citealt{hub2015}). We prioritised these events for spectroscopic classification because they exhibited a contrast between the transient and host galaxy of $\gtrsim 2$\,mag (galaxy magnitudes were taken from the Pan-STARRS DR1 catalog; \citealt{flewelling2016}). This method efficiently selects for bright SNe in faint galaxies, such as superluminous supernovae \citep{qui2011}. However, since the volumetric rate of SNe II is greater than the SLSN rate by 4 orders of magnitude, many of the SNe we classified were spectroscopically-normal SNe II. We investigate the importance of our selection effects in section \ref{sec:diss}.

Classification spectra were obtained using the FAST and Blue Channel spectrographs on the 60'' and MMT telescopes, respectively, at Fred Lawrence Whipple Observatory (FLWO), and the IMACS and LDSS spectrographs on the Magellan Baade and Clay telescopes \citep{fabricant1998,schmidt1989,dre2011,stevenson2016}. FAST spectra were reduced using a dedicated pipeline, while those from other instruments were reduced in \textsc{pyraf}. One spectrum, of PS17aio, was taken from the extended Public ESO Spectroscopic Survey of Transient Objects \citep[ePESSTO;][]{sma2015}. Redshifts were determined from host galaxy emission lines where possible; otherwise the redshift of the best-matching spectrum from SNID \citep{blondin2007} was used. All spectra will be made available via WISeREP \citep{yar2012} and the Open Supernova Catalog \citep{guillochon2017}.

\begin{figure*}
	\centering
	\includegraphics[scale=0.56,trim={2cm 0  5cm 0}]{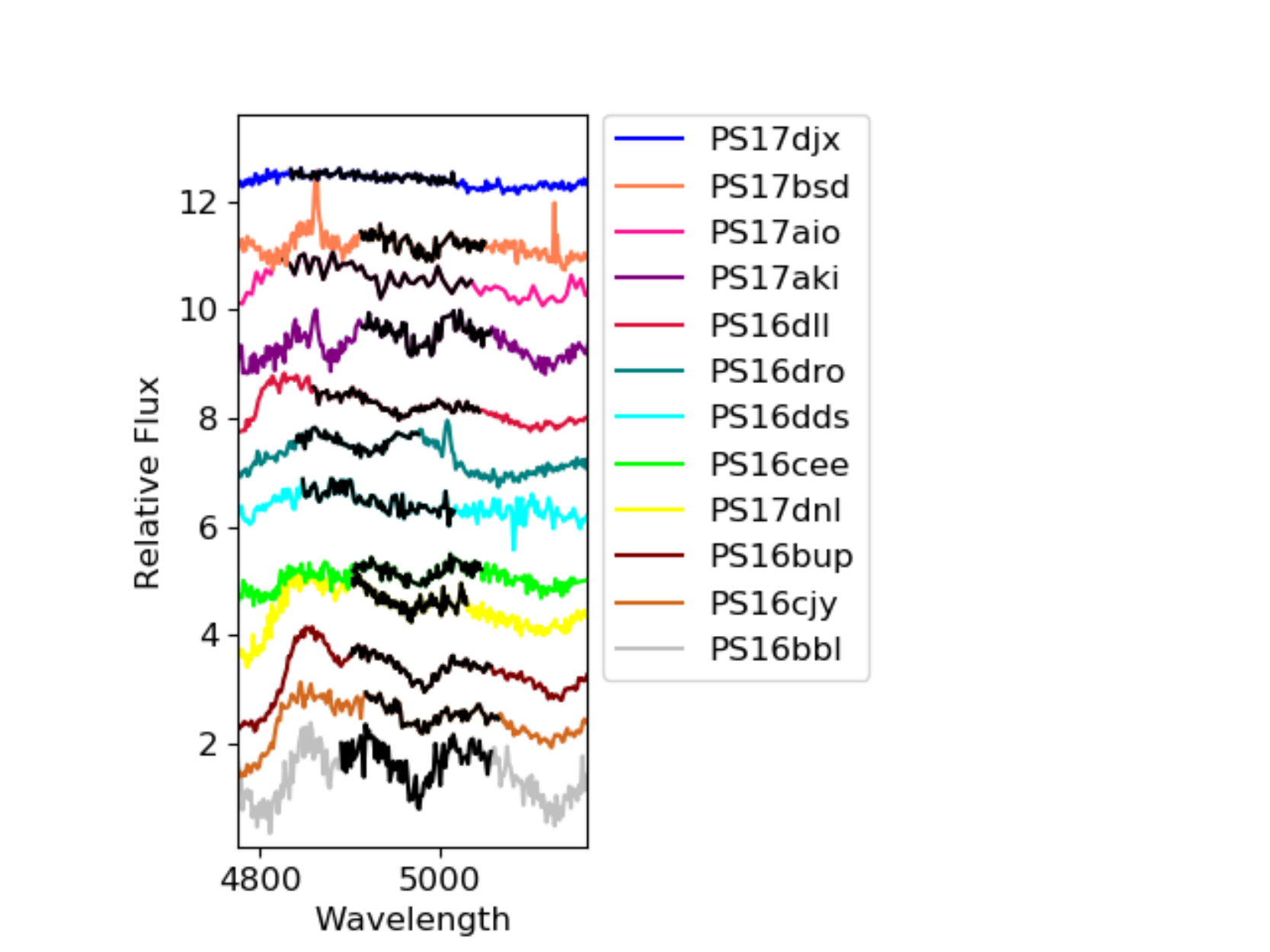}
	\includegraphics[scale=0.56,trim={0cm 0  2cm 0}]{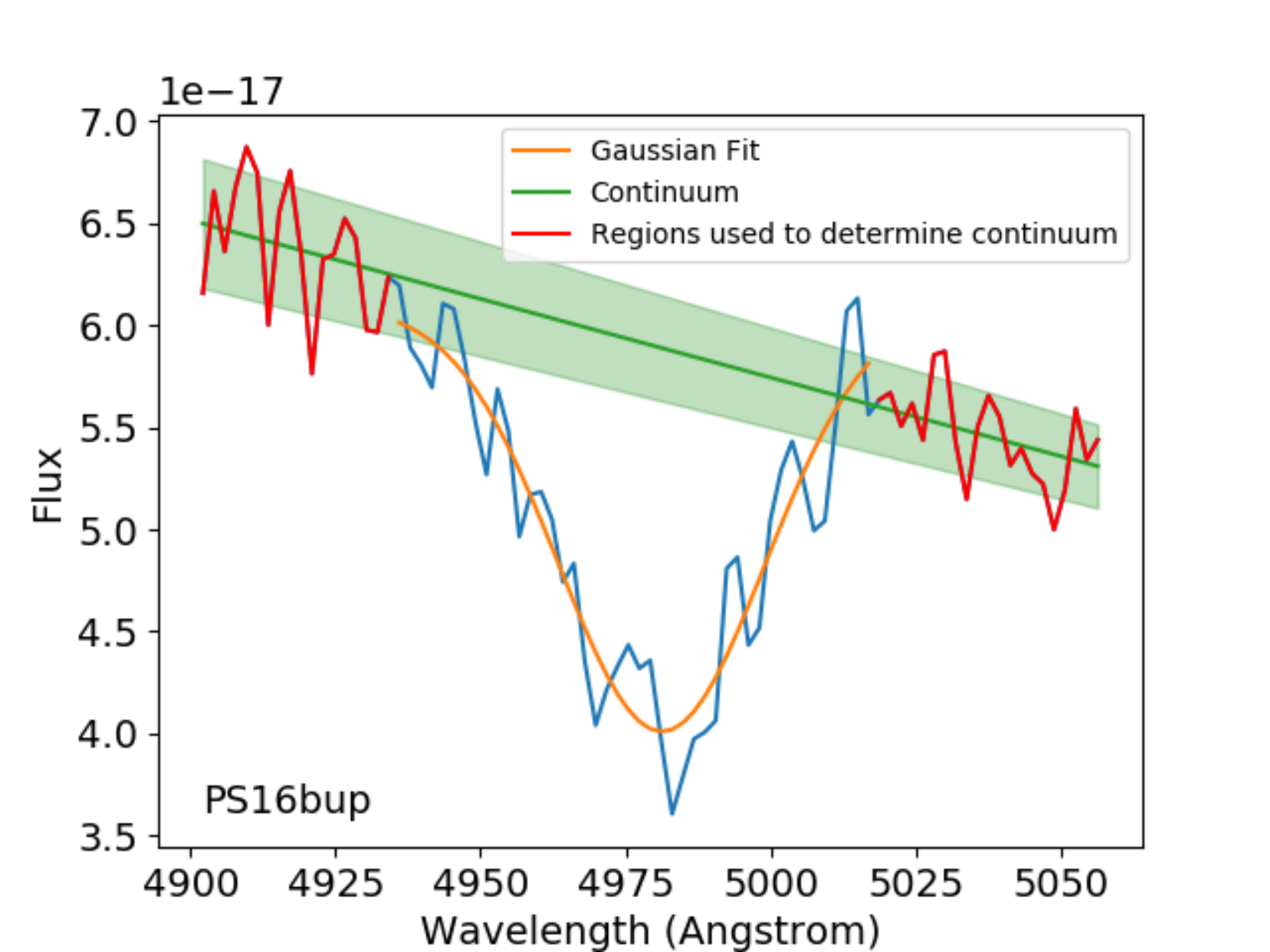}
	\caption{Left: \fe\ region for all SNe in our sample.
		Spectra are plotted in order of increasing \ion{Fe}{2} pEW, and have been normalised and shifted vertically for clarity.
		The \fe\ absorption feature and continuum region used to measure the pEW are colored black for simplified identification.
		Right: example of a Gaussian fit to the \ion{Fe}{2} line and continuum fit with a first-order polynomial used to measure the pEW.
	The upper and lower polynomials correspond to 1$\sigma$ contours around the continuum.}
	\label{f:ew}
\end{figure*}

\subsection{Pseudo-equivalent width measurements}

The equivalent width of a spectral line is defined as

\begin{equation}
	\label{EW}
	W_{\lambda} = \int(1-\frac{F_\lambda}{F_{0}})d \lambda,
\end{equation}
where $F_{\lambda}$ is the flux of the absorption feature, and $F_0$ is the flux in the continuum.
In the case of SN spectra, the true continuum is often difficult to place due to blending of many broad absorption and P Cygni features.
Instead, it is useful to define a `pseudo' equivalent width (pEW), where $F_0$ is assumed to be the peak flux on either side of the absorption line in question.

For each spectrum, we measure the pEW of the \fe\ absorption feature.
We approximate the pseudo-continuum with a linear interpolation across the absorption line, fitting the linear model to two regions surrounding the line.
The line was identified visually for each spectrum. We found that the velocity maxima of the absorption generally lay between 4925\,\AA\ (a blueshift of $\sim 5500$\,km\,s$^{-1}$) and the rest-frame wavelength of 5018\,\AA.

We then fit a Gaussian profile to the line using the least squares fit as implemented by \textsc{scipy.curve\_fit}, and integrated equation \ref{EW} taking this Gaussian fit as $F_\lambda$ and the linear fit as $F_0$.
Errors are estimated by adjusting the continuum by $\pm 1 \sigma$, where $\sigma$ is the standard deviation of the flux in the regions used to define this continuum.
Figure \ref{f:ew} demonstrates this method, and shows a close-up of the \ion{Fe}{2} line for all SNe in our sample.

Some of our spectra show no clear detection of \ion{Fe}{2} absorption features.
In this case, we place an upper limit on the pEW using the method presented in \citet{leonard2001} \citep[see also][]{graham2017}, where the $3 \sigma$ upper limit $W_{\lambda}(3\sigma)$ is defined by 

\begin{equation}
\label{eq:upperlim}
	W_{\lambda}(3\sigma) = 3 \Delta \lambda \Delta I \left(\frac{W_{line}}{\Delta \lambda \times B}\right)^{1/2}
\end{equation}
where $W_{line}$ is the width of the absorption feature, $\Delta \lambda$ is the spectral resolution (ranging from 6-18\,\AA), and $\Delta I$ is the $1 \sigma$ rms fluctuation of the normalized continuum. $B$ is the ratio of the spectral resolution to the pixel scale; $B=3-6$ for our spectra. We define $W_{line}=5018-4925=93\AA$.

\subsection{Light curves}
\label{lc_section}
The strength of spectral lines in SNe evolves over time as the ejecta expand and cool.
It is therefore important to determine the phase relative to explosion at which our spectra were obtained.
We estimate this using the light curves from PSST.
Photometry from PSST is primarily in the $r$ and $w$ filters, and occasionally in $i$ (all in the PanSTARRS system described by \citealt{ton2012}).
To calculate absolute rest-frame $r$-band magnitudes, we calculated $K$-corrections from our spectra using the s3 package presented by \citet{cinserra2018}, using cross-filter corrections where necessary. Distances were computed assuming a Planck cosmology \citep{planck2016}. Photometry was also corrected for Galactic extinction using the E(B-V) values estimated using \citet{schlaf2011}. However, no correction for internal host galaxy extinction is applied at this point in the analysis, as this parameter is more difficult to measure reliably. We will discuss the significance of this in section \ref{sec:diss}.

\begin{figure}
	\centering
	\includegraphics[scale=0.48]{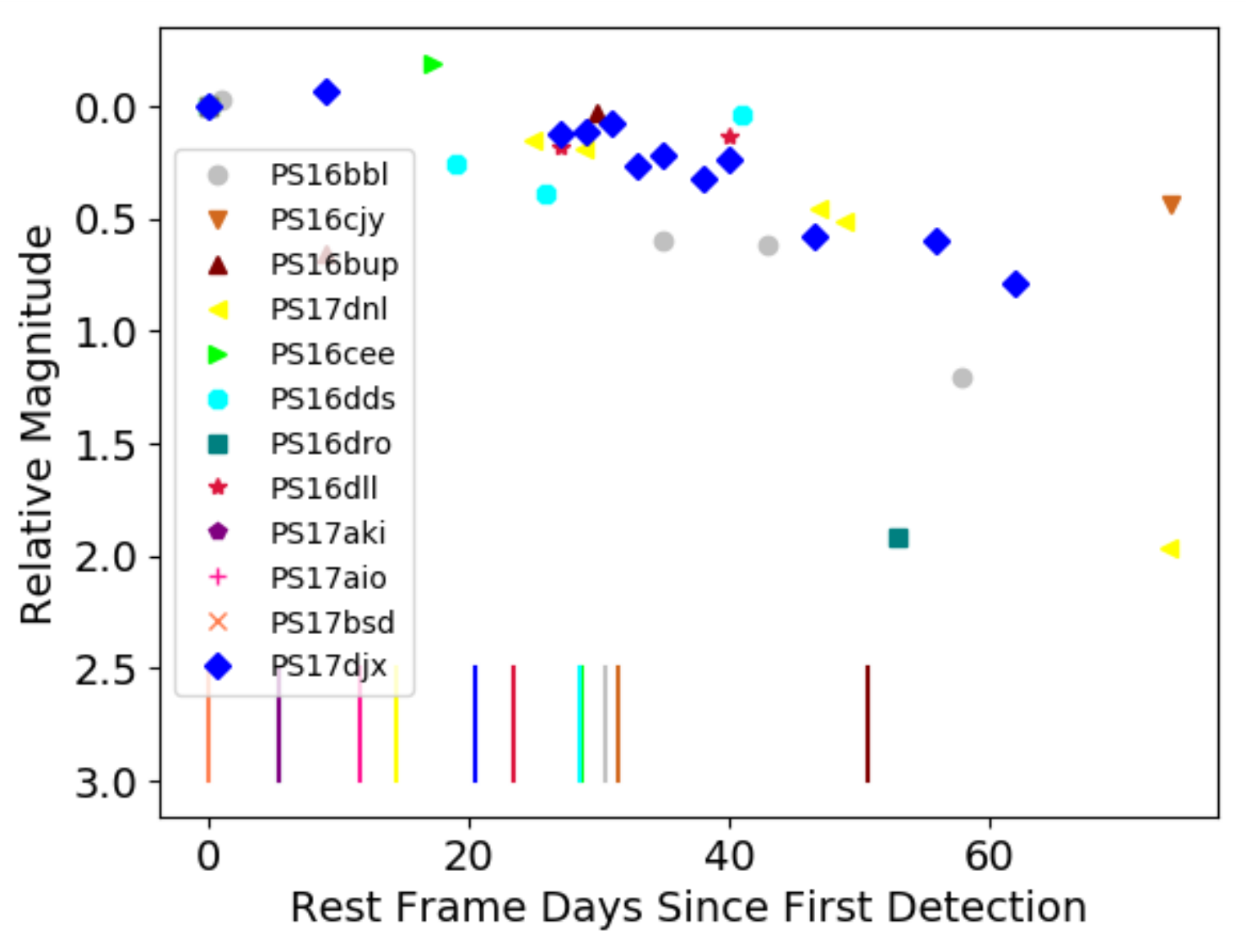}
	\caption{Light curves for all of our PSST SNe. Data have been converted to rest-frame $r$-band and shifted to match at $t=0$.
		Ticks denote dates when spectra were obtained.
		Time zero has been defined as the first data point for each light curve, and a correction for cosmological time dilation applied.
	The light curves are normalised such that the first point corresponds to 0 magnitudes. 
    SNe PS17aio, PS17bsd, and PS17aki only have one photometric data point, so each object is plotted only at $t=0$.
    }
	\label{all_lcs}
\end{figure}

The light curves are shown in Figure \ref{all_lcs}, with epochs of spectroscopy marked.
\citet{anderson2014} defined three separate epochs of SN II light curve evolution: decline from maximum, plateau, and decline on the radioactive tail.
We measure the slopes of the light curves of the SNe in our sample to ensure that we are in the plateau regime.
A linear fit to the light curves of each SN in our sample yielded a mean slope of 0.012\,mag per day, with a standard deviation of 0.016. This is similar to the mean slope during the plateau (their `s$_2$' parameter) reported by \citet{anderson2014}. This suggests that our spectra were indeed obtained during the plateau. 

We estimate the phase of the spectra, in rest frame days since explosion, and the corresponding uncertainty by assuming a simplified SN II light curve morphology with an instantaneous rise followed by a 100 day plateau and then a sharp drop.
We determine the minimum possible phase by assuming the first detected point on the PSST light curve is the date of the explosion; in this case the time of the spectrum is simply the rest-frame time since the first PSST data point.
We determine the maximum possible phase by assuming the last data point on the PSST light curve is just before the plateau phase ends; in this scenario the spectrum would correspond to a phase of 100 days minus the time from the spectrum until the end of the plateau.
The estimated time is thus defined as the mean between the upper and lower bounds, and the uncertainty is defined as the difference between the mean and these upper and lower bounds. We use this range of dates along with the light curve slopes to estimate the magnitude at the middle of the plateau.

\subsection{Comparison sample}
\label{compare}

To understand how the properties of SNe II in dwarf galaxies differ from those in massive galaxies, we construct a comparison sample of literature SNe II using the Open Supernova Catalog (OSC; \citealt{guillochon2017}).
We first filtered supernovae by type (II or IIP), and then required that each SN have at least 5 spectra available in the catalog, in order to select only well-studied events.
These criteria narrowed the literature sample down to 24 SNe.
We were specifically interested in spectra obtained during the middle of the plateau phase (i.e.~around 50 days after light curve maximum).
We visually inspected all light curves to ensure that the automatically-derived date of maximum light from the OSC was accurate, and corrected this when necessary.
The spectrum that was closest to 50 days was then used for analysis.

Many of the literature SNe were observed only in the Johnson-Cousins photometric bands. For a fair comparison with our sample, we applied a cross-filter $K$-correction using S3 \citep{cinserra2018} to convert to rest-frame $r$-band in the PanSTARRS system. We again corrected for Milky Way extinction only. Because most of these SNe are relatively nearby, we used redshift-independent distances from the Nasa Extragalactic Database (NED). In most cases we used the median of the reported distance estimates. Some objects had a SN distance from the expanding photosphere method \citep{schmidt1994,poznanski2009,rod2014,tak2015} that differed significantly from the median distance, in which case we selected the SN distance.

\subsection{Statistical Tests}
\label{stat_tests}

The parameters of interest in this study are the pEW of \fe, the absolute magnitude of the SN at the middle of the plateau, and the absolute magnitude of the host galaxy. The latter is a proxy for the mass of a galaxy and hence its metallicity \citep[e.g.][]{tremonti2004,kewley2008}. We wish to test whether these parameters are significantly different between our PSST sample and the literature control sample.

In order to do this, we employ a 2 sample Kolmogorov-Smirnoff (KS) test implemented in \textsc{scipy.ks\_2samp}.
The 2 sample KS test enables us to test the null hypothesis that the two samples are drawn from the same distribution.

In the case of the \fe\ line, there were 3 SNe for which we were only able to establish upper limits on the pEW.
Therefore, a 2 sample KS test is not appropriate, because a KS test does not account for upper limits.
Instead, we use a Wilcoxon rank-sum test, which, in the case where no upper limits are present, reduces to a 2 sample KS test \citep[see][for the use of this test in a similar context]{lun2014}.

\section{Results}
\label{sec:results}

\begin{figure*}
	\centering
	\includegraphics[scale=0.6]{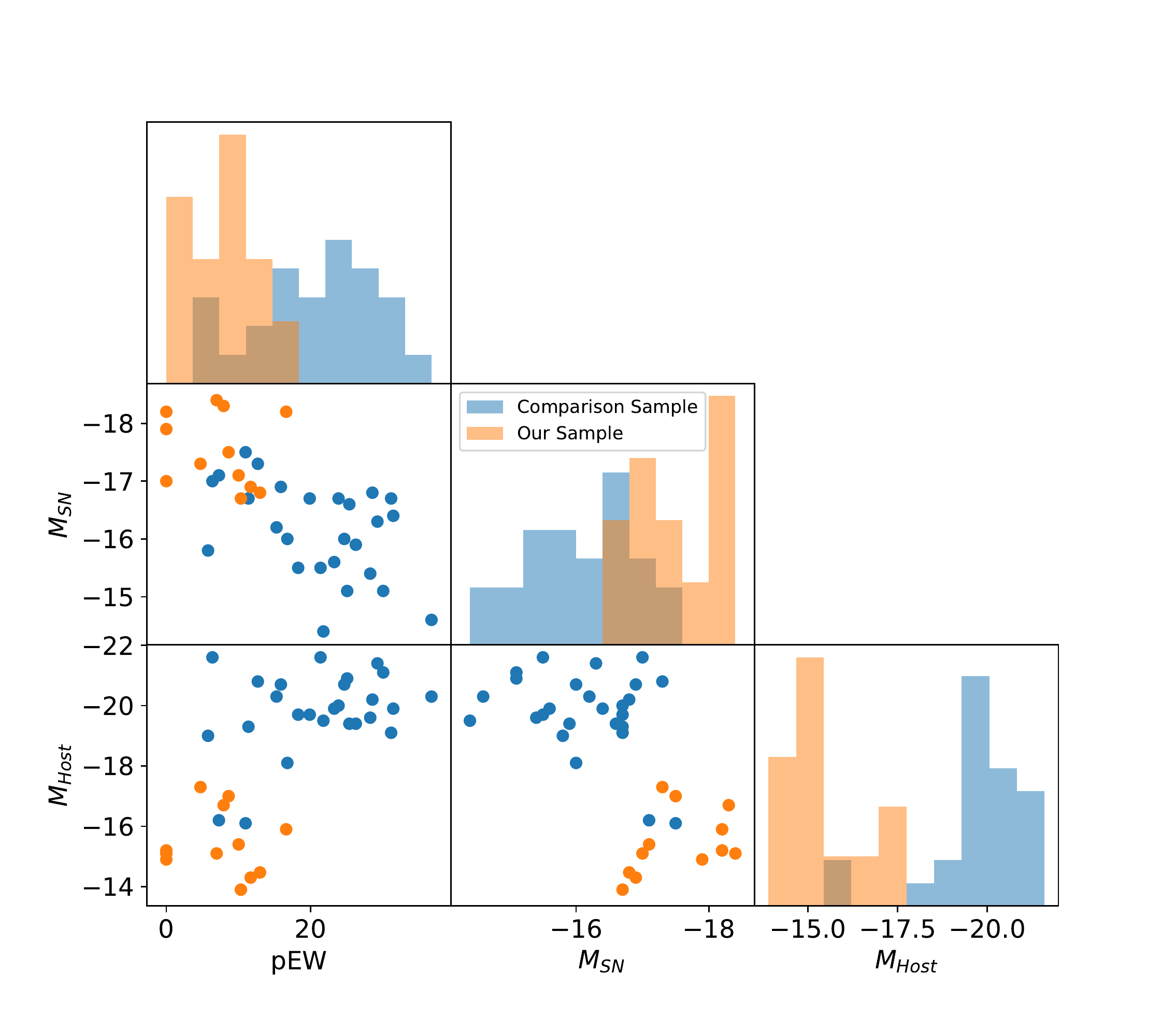}
    \includegraphics[scale = 0.25]{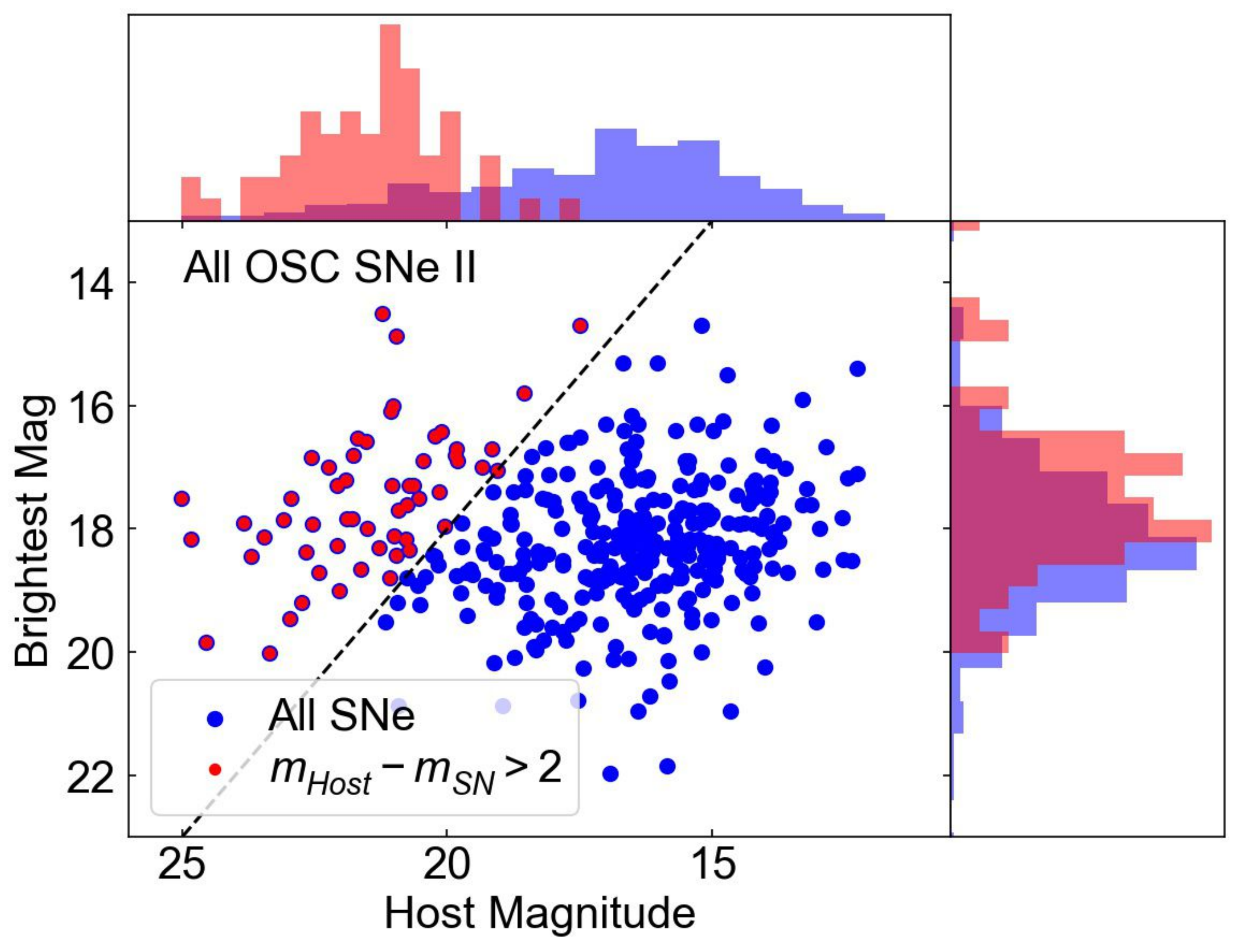}
    \includegraphics[scale=0.25]{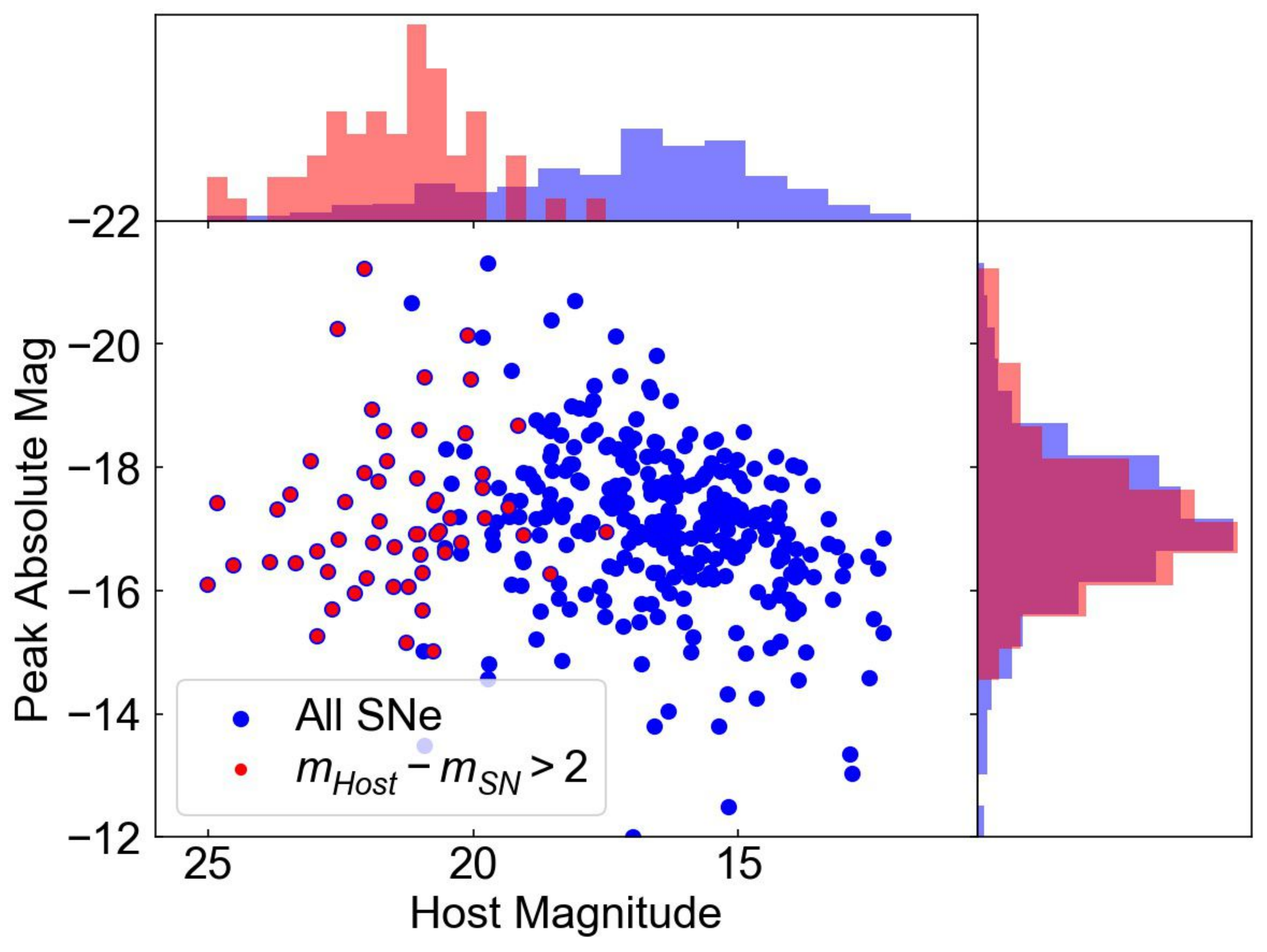}
	\caption{Top: Corner plot of \fe\ pEW, $M_{SN}$, and $M_{host}$. The normalized histograms illustrate the differences in these properties measured in our sample and in the catalog sample. The trend found in pEW vs $M_{host}$ indicates that \fe\ in SNe II is a good tracer of host metallicity. The SNe in faint galaxies also appear to be brighter, though we caution that host galaxy extinction has not been applied here (see section \ref{sec:ext}).
    Bottom: Quantifying our selection effects using the Open Supernova Catalog. Blue points show all SNe II, red points show those that satisfy our selection criteria. Left: apparent magnitudes; right: absolute magnitudes. The bias in SN absolute magnitude introduced by our selection effects (section \ref{sec:select}) is much smaller than the size of the SN metallicity-magnitude correlation.
    }
	\label{distributions}
\end{figure*}

Figure \ref{distributions} shows a corner plot with the distributions of SN and host absolute magnitudes and \fe\ pEW, for both the PSST sample and comparison sample. We also list these measurements in Table \ref{sample_properties}.
We describe the statistical results for each set of measurements below.

\subsection{Galaxy Properties}
We first compare the photometric properties of the host galaxies.
We find the mean $M_{host}$ in our sample is $-19.8$ with standard deviation 1.3, while the mean brightness of host galaxies within the comparison sample is $-15.4$ with standard deviation 1.0.
Using a two sample KS test as described in Section \ref{stat_tests}, we reject the null hypothesis that the brightness distributions of the host galaxies in each sample come from the same distribution with a p-value of $3\times10^{-7}$. We therefore confirm that our SNe are in a significantly less luminous galaxy population than the control sample, and hence should probe SN II properties at lower metallicities.

\subsection{SN Spectroscopic Properties}
Comparing the pEW of the \fe\ absorption features of the two samples, we calculate a mean pEW in our sample of 7.5 \AA\ with standard deviation 5.2 \AA, and a mean pEW in the comparison sample of 21.1 \AA\ with standard deviation 8.3\AA.
We apply the Wilcoxon rank-sum test due to the presence of upper limits in the pEW measurements in our sample, as described in section \ref{stat_tests}.
Using this test, we reject the null hypothesis that the \fe\ pEWs from the two samples are drawn from the same distribution with a p-value of $10^{-4}$. Combining this with our knowledge that the host galaxies in our sample are significantly fainter and therefore likely more metal-poor than the control sample, this confirms that SNe II at low metallicity show weaker \fe\ lines.

\subsection{SN Photometric Properties}
The mean $M_{SN}$ in our sample (estimated at the t=50 days epoch following the method in Section \ref{lc_section}) is $-17.5$, with standard deviation 0.6. For the comparison sample, the mean $M_{SN}$ is $-16.1$ with standard deviation 0.8.
Utilizing the KS test as before, we reject the null hypothesis that the SNe in each sample come from the same brightness distribution with a p-value of $3\times10^{-4}$. While at first sight the luminosity differences between the samples is striking, there are important caveats due to selection effects and host galaxy extinction as we discuss below.

\subsubsection{Possible selection effects in SN luminosity}

At least some of the difference between the two samples is likely a consequence of our targeted search for SNe that outshine their host galaxies, which clearly favours brighter SNe. 
To quantify our selection effects, we make use of the Open Supernova Catalog. We downloaded all SNe II in the catalog, and in the lower panels of Figure \ref{distributions} we plot SN peak magnitudes against host magnitudes, indicating any SNe that would have passed our selection cuts (section \ref{sec:select}).
The median apparent magnitude of all SNe before data cuts is $18.2\pm1.2$, and after our data cuts, the median apparent magnitude changes to $17.7\pm1.3$. 
Applying the same methods to absolute magnitudes instead of apparent magnitudes, the medians before and after cuts are $-17.1\pm1.3$ and $-16.9\pm1.5$. 

Therefore, although our data cuts unsurprisingly introduce a non-negligible bias, of up to $\sim 0.5$\,mag, in the apparent magnitudes of selected events, this discrepancy disappears into the scatter in $M_{SN}$ vs $M_{host}$ when corrected to an absolute magnitude scale. This assumes that all SNe passing the cuts are equally likely to be classified; a somewhat larger selection effect could result if the likelihood of classification increases with SN-host contrast. However, this experiment indicates that it is unlikely for selection effects alone to account for the entire 1.4 mag difference between the PSST sample and the comparison sample.

\subsubsection{The impact of host galaxy extinction}
\label{sec:ext}

Another important factor to consider is dust extinction within the host galaxies. We have not corrected the SNe in either the low-metallicity or comparison sample for internal extinction. More massive galaxies generally exhibit higher extinction. \citet{garn2010} provide an empirical relationship between stellar mass and extinction, finding that an extinction in H$\alpha$ (similar to the wavelength of $r$-band) of 1.5 magnitudes (i.e.~the difference in brightness between our two samples) is typical of galaxies with stellar masses $\gtrsim 5\times 10^{10}$\M. Many of the literature SNe are indeed in galaxies within this mass range.

To test whether extinction alone can account for the sample luminosity differences, we estimate host galaxy extinctions, $A_r$, from $r$-band galaxy luminosities using the relation of \citet{garn2010} and the SDSS mass-to-light ratios from \citet{kauffmann2003} ($M_*/L_r\approx3$). These values are listed in Table 1. These relations break down for the dwarf galaxies, and unfortunately none of our sample showed strong galaxy emission lines from which we could reliably calculate a Balmer decrement. Therefore we assume a modest $A_r = 0.3$\,mag (where the \citealt{garn2010} relation flattens out). If we include these extinction estimates, we find a mean sample magnitude of $-17.8$ mag compared to a mean literature magnitude $-17.3$ mag. In this case, the KS test gives a p-value 0.13, i.e.~the luminosity difference is no longer statistically significant.

\section{Discussion}
\label{sec:diss}

\subsection{Metallicity}

Our pEW results clearly support previous findings from \citet{dessart2014,anderson2016,taddia2016a,gutierrez2018}. Our SNe are in faint galaxies (absolute $M_r > -17$), which are expected to be of low metallicity. Using the empirical calibration from \citet{arcavi2010}, this corresponds to metallicities $\lesssim 0.2$\,\Z\ (though likely with significant scatter). All the SNe II in these galaxies had \fe\ pEW $< 15$\,\AA, well below typical values for SNe II in massive galaxies. Thus we confirm that the pEW of the \fe\ line is an efficient way to select low-metallicity SNe II.

\begin{figure}
	\centering
	\includegraphics[scale=0.48]{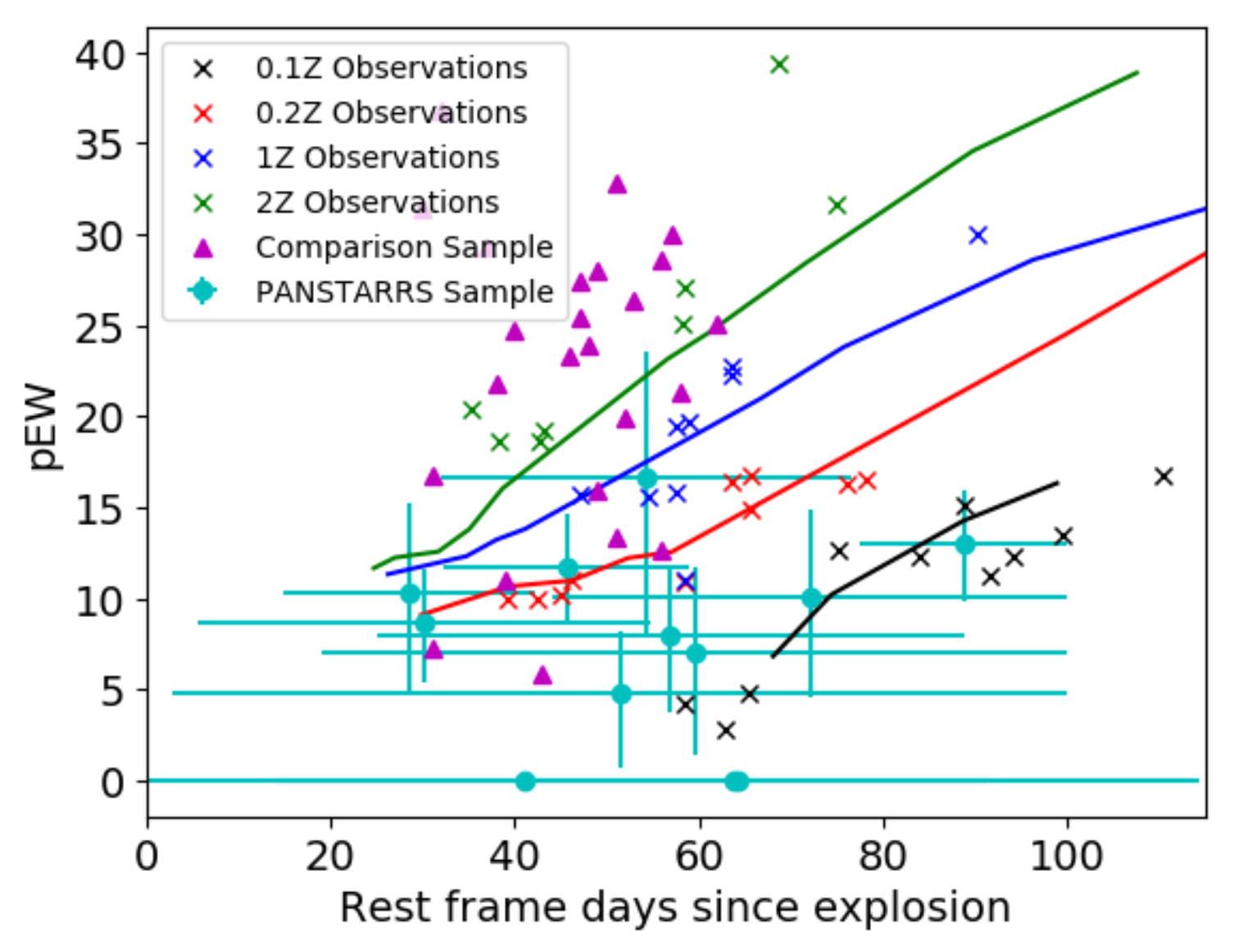}
	\caption{Top: evolution of the pEW of \fe\ over time for our sample (cyan circles) and the catalog sample (magenta triangles). We also plot spectral models from \citet{dessart2014}, at metallicities between 0.1--2\,Z$_\odot$, as solid lines and the PTF sample from \citet{taddia2016a} as crosses.
    }
	\label{pew_time}
\end{figure}

In Figure \ref{pew_time}, we compare our measurements to the spectral models from \citet{dessart2014} and the unbiased sample from \citet{taddia2016a}. Almost all of our pEW measurements are consistent with metallicities $\lesssim 0.4$\,\Z, with several matching models at 0.1\,\Z. This is consistent with the metallicities estimated from their host luminosities \citep{arcavi2010}. The control sample of literature SNe are mainly consistent with Solar metallicity or greater.

\subsection{Luminosity}

Interestingly, we find that the SNe in our sample may be more luminous on average than the literature sample, suggesting a possible relationship between metallicity and SN II luminosity. This is clear in the $M_{SN}$ vs pEW panel of Figure \ref{distributions}. 
However, the significance of any such relation is highly sensitive to assumptions about extinction in the SN host galaxies.

Several SNe in our sample are brighter than any SNe in the control sample, at least indicating that luminous SNe II do occur at low metallicity. \citet{taddia2016a} also observed more luminous SNe II at lower metallicity, finding that their low-metallicity events (black and red points in Figure \ref{pew_time}) were brighter by 0.7\,mag than the high-metallicity events (blue and green points). Thus the size of the effect they measured is comparable to our findings here.

More recently, \citet{gutierrez2018} found a correlation between the \fe\ pEW and the gradient of the light curve during the plateau phase. Since our luminosity measurements were made as close as possible to the middle of the plateau, a slower decline would result in a brighter measurement at the mid-point. Therefore a shallower light curve at low metallicity would also help to explain our findings. However, we note that our measured slopes were steeper on average than the sample from \citet{gutierrez2018}, and more consistent with the \citet{anderson2016} sample. On the other hand, the light curves obtained from PSST are sparsely sampled, so we may be fitting to points outside of the plateau phase. Therefore, we cannot robustly discriminate between the interpretations of \citet{taddia2016a} and \citet{gutierrez2018}.

\section{Conclusions}
\label{sec:conc}
In this work, we presented a new sample of 12 SNe II that occurred in low-luminosity ($M\gtrsim -17$) host galaxies.
We measured the pseudo-equivalent width of the \fe\ absorption feature in the SN spectra, finding that the line was significantly weaker than in SNe II from a literature control sample, which were primarily in massive host galaxies. This is consistent with expectations that the low-mass galaxies are metal-poor, and confirms previous results from \citet{anderson2016,taddia2016a,gutierrez2018}, and theoretical predictions by \citet{dessart2014}, indicating that \fe\ is a useful proxy for metallicity in SNe II.

We also found evidence for a possible correlation between SN and host luminosity, with more luminous SNe occurring in low-luminosity galaxies. While extinction in the host galaxies clearly plays an important role, it appears that there could be a luminosity difference of up to $\sim 0.5$\,mag between type II SNe at low and high metallicity. Future studies can test our results by better estimating internal extinction, or observing SNe in the infrared where extinction effects are minor. Larger samples will also be needed to determine statistical significance. Simulating our search criteria on a larger sample from the Open Supernova Catalog suggested that selection effects are minor, indicating that any difference is most likely physical. 

Our results add to the growing importance of metallicity as a determining factor in the diverse outcomes of massive star evolution. In the future, larger and more homogeneous samples of SNe II, from surveys such as ZTF and LSST, will improve our understanding of the physical effects underpinning this diversity.

\acknowledgements
We thank an anonymous referee for many helpful comments that greatly improved this manuscript.
M.N.~is supported by a Royal Astronomical Society Research Fellowship.
The Berger Time-Domain Group at Harvard is supported in part by the NSF under grant AST-1714498 and by NASA under grant NNX15AE50G. P.K.B.~acknowledges NSFGRP Grant No.~DGE1144152.
This paper includes data gathered with the 6.5 meter
Magellan Telescopes located at Las Campanas Observatory,
Chile.
Some observations
reported here were obtained at the MMT Observatory, a joint
facility of the Smithsonian Institution and the University of
Arizona.
This paper uses data products
produced by the OIR Telescope Data Center, supported by the
Smithsonian Astrophysical Observatory. 
This research has made use of the NASA/IPAC Extragalactic Database (NED) which is operated by the Jet Propulsion Laboratory, California Institute of Technology, under contract with the National Aeronautics and Space Administration.

\begin{table*}
	\centering
	\caption{A table of the measured properties of the SNe used in this study.}
	\label{sample_properties}
	\begin{tabular}{llllll}
		Name    & $z$      & \ion{Fe}{2} pEW   & $M_{SN}^*$ & $M_{host}^*$ & $A_{r,host}^\dagger$\\
		\hline
		PS16bbl & 0.055 & $16.6^{7.0}_{8.8}$ & $-18.2$    & $-15.9$ & 0.3   \\
		PS16bup & 0.038 & $13.0^{2.9}_{3.2}$ & $-16.8$    & $-14.7$  & 0.3  \\
		PS16cee & 0.0297 & $10.0^{4.8}_{5.5}$ & $-17.1$    & $-15.4$  & 0.3  \\
		PS16cjy & 0.023 & $11.7^{2.9}_{3.2}$ & $-16.9$    & $-14.3$   & 0.3 \\
		PS16dds & 0.104 & $< 2.9$ & $-18.3$    & $-16.7$             &  0.3  \\
		PS16dll & 0.067 & $8.0^{3.9}_{4.3}$  & $-17.5$    & $-17.0$   & 0.3  \\
		PS16dro & 0.0437 & $8.7^{3.0}_{3.2}$  & $-18.4$    & $-15.1$  & 0.3   \\
		PS17aki & 0.075 & $7.0^{4.8}_{5.6}$  & $-17.3$   & $-17.3$    & 0.3  \\
		PS17bsd & 0.028 & $4.7^{3.5}_{4.1}$  & $-16.7$   & $-13.9$   &  0.3  \\
		PS17dnl & 0.039 & $10.4^{4.9}_{5.6}$ & $-17.9$   & $-14.9$    & 0.3  \\
		PS17aio & 0.048 & $< 3.9$ & $-17.0$    &  $-15.1$             & 0.3  \\
		PS17djx & 0.045 & $< 1.7$ & $-18.2$    &  $-15.2$            &  0.3  \\
		\hline
		\hline
		SN1969L  & 0.00253  & $11.4^{2.4}_{2.6}$	& $-16.7$ & $-19.3$ & 1.0  \\
        SN1990E  & 0.00429  & $5.8^{1.2}_{1.2}$		& $-15.8$ & $-19.0$ & 0.9  \\
		SN1992H  & 0.00457  & $12.7^{1.0}_{1.0}$ 	& $-17.3$ & $-20.8$ & 1.5  \\
		SN1999br & 0.003201 & $21.8^{2.4}_{2.7}$	& $-14.4$ & $-19.5$ & 1.1  \\ 
		SN1999em & 0.002392 & $25.4^{0.3}_{0.3}$	& $-16.6$ & $-19.4$ & 1.0  \\
		SN1999gi & 0.001975 & $16.8^{1.3}_{1.4}$	& $-16.0$ & $-18.1$ & 0.6  \\
		SN2001X  & 0.004937 & $28.6^{0.6}_{0.6}$	& $-16.8$ & $-20.2$ & 1.3  \\
		SN2002gd & 0.00774  & $23.3^{0.8}_{0.8}$	& $-15.6$ & $-19.9$ & 1.2  \\
		SN2003Z  & 0.00629  & $25.1^{0.4}_{0.4}$	& $-15.1$ & $-20.9$ & 1.5  \\
		SN2003gd & 0.002192 & $24.7^{3.3}_{3.8}$	& $-16.0$ & $-20.7$ & 1.4  \\
		SN2003hl & 0.00825  & $29.3^{2.9}_{3.2}$	& $-16.3$ & $-21.4$ & 1.7  \\
		SN2004A  & 0.00459  & $31.5^{0.5}_{0.5}$	& $-16.4$ & $-19.9$ & 1.2  \\
		SN2004dj & 0.000456 & $26.3^{0.4}_{0.4}$	& $-15.9$ & $-19.4$ & 1.0  \\
		SN2004du & 0.016762 & $7.3^{3.5}_{4.0}$		& $-17.1$ & $-16.2$ & 0.3  \\
		SN2004et & 0.000909 & $19.9^{0.9}_{0.9}$	& $-16.7$ & $-19.7$ & 1.1  \\
		SN2005ay & 0.002699 & $23.9^{0.6}_{0.6}$	& $-16.7$ & $-20.0$ & 1.2  \\
		SN2005cs & 0.00137  & $30.1^{0.4}_{0.5}$	& $-15.1$ & $-21.1$ & 1.6  \\
		SN2006bp & 0.003509 & $18.3^{0.7}_{0.7}$	& $-15.5$ & $-19.7$ & 1.1  \\
		SN2007od & 0.00586  & $11.0^{0.9}_{1.0}$	& $-17.5$ & $-16.1$ & 0.3  \\
		SN2009N  & 0.003449 & $28.3^{0.4}_{0.4}$	& $-15.4$ & $-19.6$ & 1.1  \\
		SN2009ib  & 0.00448 & $21.4^{1.2}_{1.3}$	& $-15.5$ & $-21.6$ & 1.7  \\
		SN2010aj  & 0.0212  & $6.4^{2.5}_{2.8}$ 	& $-17.0$ & $-21.6$ & 1.7  \\
		SN2012aw & 0.002595 & $31.2^{1.0}_{1.0}$	& $-16.7$ & $-19.1$ & 0.9  \\
		SN2012ec & 0.004693 & $15.3^{0.7}_{0.7}$	& $-16.2$ & $-20.3$ & 1.3  \\
		SN2013am & 0.002692 & $36.8^{3.4}_{4.3}$	& $-14.6$ & $-20.3$ & 1.3  \\
		SN2013ej & 0.002192 & $15.9^{0.8}_{0.8}$	& $-16.9$ & $-20.7$ & 1.4  \\		\hline \\
	\end{tabular}
    
        $^*$ Magnitudes listed are all in r-band, after correcting for distance, Milky Way extinction and $K$-corrections (and filter corrections where necessary). Internal host galaxy extinction has not been applied here.
        
        $^\dagger$ Estimated extinction from the host galaxy in $r$-band, calculated using the absolute magnitudes and the mass-to-light ratios from \citet{kauffmann2003} and the mass-extinction relation from \citet{garn2010}. These relations break down at the faint end, where we assume a default extinction of 0.3\,mag.
        
\end{table*}

\end{document}